\documentclass[pre,preprint,superscriptaddress,amsmath,amssymb,aps,longbibliography]{revtex4-2}

\usepackage{graphicx}          
\usepackage{booktabs}          
\usepackage{dcolumn}           
\usepackage{bm}                
\usepackage{hyperref}          
\hypersetup{colorlinks=true, linkcolor=blue, citecolor=blue, urlcolor=blue}
\usepackage{cleveref}          

\graphicspath{{figures/}}

\begin{document}
\title{Resolving the Gibbs Paradox through Local Free Space and Collision Potential}

\author{Chuang Li}
\email{lichuang@alumni.itp.ac.cn}
\affiliation{College of Mechanical and Electrical Engineering, Wuyi University, Nanping 354300, China}


\date{\today}
	
\begin{abstract}
	The Gibbs paradox is conventionally resolved by invoking the indistinguishability of identical particles and the associated $N!$ correction, while the standard entropy of mixing remains independent of molecular properties. Here, we develop a classical framework that does not rely on quantum indistinguishability. We treat molecules initially occupying different spatial regions as distinguishable and describe the accessible volume of each molecule by its instantaneous local free space. We further account for the non-negligible time-averaged potential energy associated with molecular collisions and rederive the entropy increment of gas mixing within the canonical ensemble. The resulting expression depends explicitly on molecular properties, including molecular mass, effective radius, and characteristic collision time, and therefore varies among different gases while yielding zero entropy increment for identical gases naturally. This framework provides a novel classical resolution of the Gibbs paradox by connecting the entropy of mixing to molecular properties rather than to an additional quantum assumption or some new conceptual explanation.
\end{abstract}
	


\maketitle
	
\section{\label{sec:intro}Introduction}
The Gibbs paradox is a longstanding problem at the interface of thermodynamics and statistical mechanics because it exposes a tension between microscopic state counting and the macroscopic entropy of mixing \cite{StatMech:Books,StatMech:BooksAdv,Darrigol:2018,LinSK-webpage:2009}. Consider two ideal gases at equal temperature and pressure, initially separated by a partition. Let $V_i$ and $N_i$ denote the volume and number of molecules in compartment $i$, respectively, with $V=V_1+V_2$ and $N=N_1+N_2$. After removal of the partition, the gases mix and approach a uniform equilibrium state. In the conventional description, the entropy increment is
\begin{equation}
	\Delta S=S_m-\sum\limits_{i=1}^2 S_i=-k\sum\limits_{i=1}^2N_i\ln x_i,
	\label{MixingEntropyIncreaseSimple}
\end{equation}
where $S_i$ is the entropy of each component before mixing, $S_m$ is the entropy after mixing, $x_i=N_i/N=V_i/V$, and $k$ is the Boltzmann constant.

Equation \ref{MixingEntropyIncreaseSimple} leads to two closely related conceptual difficulties. First, if it is applied directly to two portions of the same gas, it predicts a nonzero entropy increment even though the initial and final macroscopic equilibrium states are thermodynamically equivalent. The standard statistical-mechanical treatment restores extensivity by accounting for particle permutations through the $N!$ factor, commonly associated with the indistinguishability of identical particles. This aspect of the paradox has been extensively discussed and is generally regarded as resolved within conventional thermodynamics and statistical mechanics \cite{Jaynes:1992,Darrigol:2018}.

A second difficulty concerns mixtures of different gases. In the conventional ideal-gas expression, the entropy increment depends on particle numbers and composition but not on intrinsic molecular properties. Consequently, as two gases are made progressively more similar, the predicted mixing entropy remains finite until the components are treated as identical, at which point it becomes zero. This discontinuity is often referred to as the ``true'' or second Gibbs paradox. The historical development and the range of proposed resolutions have been reviewed extensively \cite{Darrigol:2018}, with additional literature collected in \cite{LinSK-webpage:2009}.

Here, we examine this second problem from a classical statistical-mechanical perspective. Our starting point is that molecules initially occupying spatially separated regions can be distinguished by their spatial origin within a classical description; quantum indistinguishability is therefore not introduced as the physical mechanism that determines the entropy change. This choice, however, requires a different account of extensivity. Following the local-free-space perspective proposed in \cite{Guo:2020,Guo:2021}, we assume that the instantaneous spatial region available for the free motion of an individual molecule is local and scales with the volume per molecule rather than with the entire macroscopic volume.

Local-free-space perspective alone yields no entropy increase upon mixing, which is difficult to reconcile with irreversible diffusion between different gases. We therefore consider a second physical ingredient: intermolecular collision potential. Although the interaction potential is appreciable only during the short collision interval, collisions occur repeatedly and provide the microscopic mechanism by which a gas approaches equilibrium. We consequently retain the time-averaged collision potential energy rather than setting the intermolecular potential identically to zero at all times.

Combining local-free-space perspective with collision potential energy, we use the canonical ensemble to rederive the entropy increment of gas mixing. The resulting expression contains molecular mass, effective molecular radius, characteristic collision time, and thermodynamic variables. The mixing entropy therefore becomes dependent on the physical properties of the gases and approaches zero when the two components become identical. This formulation provides a classical, physically parameterized route to the Gibbs paradox without requiring quantum indistinguishability as the mechanism of its resolution.

\section{\label{sec:model}Model and Methods}

This paper adopts the canonical ensemble method as its theoretical framework which is explained in many textbooks\cite{StatMech:Books,StatMech:BooksAdv} and will not be repeated here. However, the canonical ensemble method requires the consideration of the following crucial theoretical assumptions; otherwise, the conclusions of this paper cannot be obtained.

\subsection{Reconsidering the $N!$ correction in a classical description}
Our analysis begins by separating the classical description used here from the conventional quantum-mechanical indistinguishability argument. For macroscopic gases occupying spatially distinct regions before mixing, the molecules can be labeled by their initial spatial origin. Within this classical description, permutation of labels need not be removed by invoking quantum indistinguishability. Related arguments concerning the distinguishability of identical particles in statistical descriptions have been discussed previously \cite{Dieks:2010,Versteegh-Dieks:2011,Dieks:2018}.

The usual $N!$ correction is closely connected to the extensivity of the entropy. Its thermodynamic-limit treatment typically uses Stirling's approximation,
\begin{equation}
	\ln N! \approx N\ln N-N,
	\label{FactorialFormula1}
\end{equation}
whereas retaining the next term gives
\begin{equation}
	\ln N! \approx N\ln N-N+\frac{1}{2}\ln(2\pi N).
	\label{FactorialFormula2}
\end{equation}
The subextensive logarithmic term illustrates that exact finite-$N$ extensivity is not obtained simply by truncating the factorial contribution. This finite-size issue has also been discussed in the context of the Gibbs paradox \cite{Casper-Freier:1973}. Rather than using the $N!$ correction as the physical origin of extensivity, we therefore ask whether extensivity can emerge from a different specification of the spatial phase space accessible to each molecule.

\subsection{Local free space as the molecular spatial volume}
The key spatial assumption is that the instantaneous free-motion volume available to an individual molecule is local. Because each molecule undergoes repeated collisions and has a finite mean free path, it does not explore the entire macroscopic volume independently at each instant. In the local-free-space perspective, the characteristic spatial volume per molecule scales as $V/N$ rather than $V$ \cite{Guo:2020,Guo:2021}. This choice provides a direct route to extensive entropy without attributing extensivity to the combinatorial removal of particle permutations.

Local-free-space perspective by itself, however, predicts no entropy increment when two gases mix. Such a result captures the identical-gas limit but does not account for the irreversible mixing of distinct gases. A further physical contribution is therefore required if the statistical description is to distinguish mixtures according to molecular properties.

\subsection{Collision potential as a contribution to mixing entropy}
In the ideal-gas approximation, intermolecular potential energy is normally neglected except during collisions. Here, we retain the collision contribution in a time-averaged description. Collisions are short-lived, but their frequency is high, and they constitute the microscopic process through which momentum and energy are exchanged and equilibrium is established. The average collision potential can therefore contribute to the canonical partition function even when the potential vanishes during most of the free-flight interval.

Figure 1 schematically illustrates the collision potential energies for molecules of type 1, type 2, and unlike molecular pairs. The relevant amplitudes and collision durations can differ between molecular species and between like and unlike collisions. These differences provide a molecular-property dependence that is absent from the conventional ideal-gas mixing expression.

\begin{figure}[htbp]
	\centering
	\includegraphics[width=320pt]{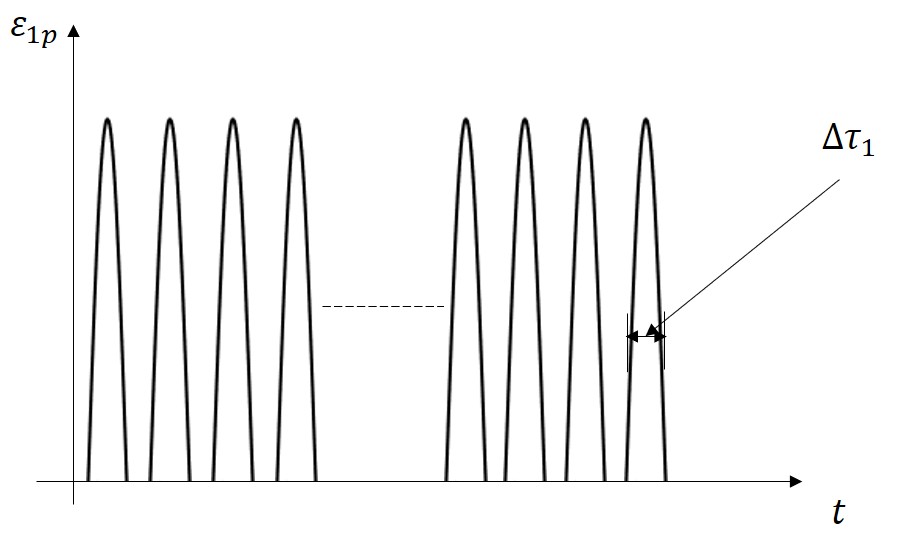}
	\caption{Schematic time dependence of the collision potential energy ${\varepsilon}_{1p}$ for molecules of type 1.}
	\vspace{5pt}
	\centering
	\includegraphics[width=320pt]{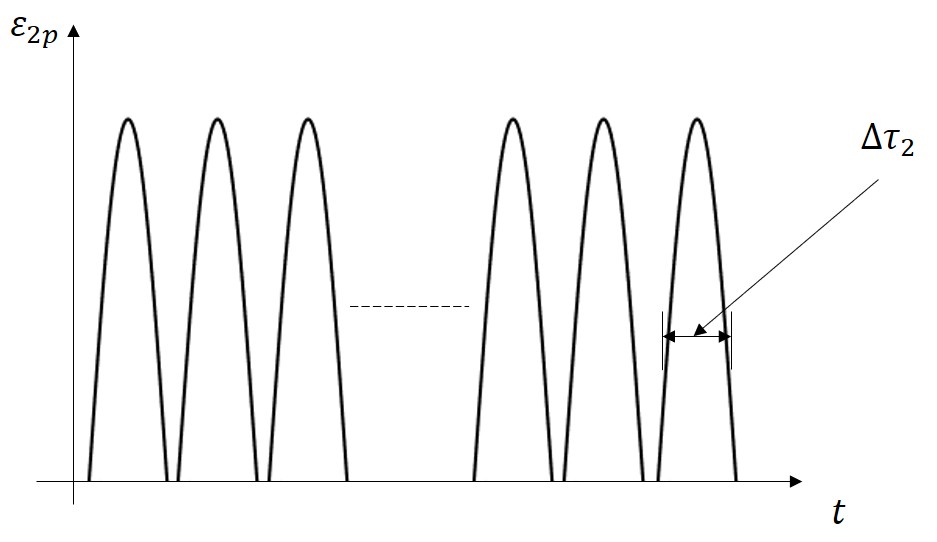}
	\caption{Schematic time dependence of the collision potential energy ${\varepsilon}_{2p}$ for molecules of type 2.}
	\vspace{5pt}
	\centering
	\includegraphics[width=320pt]{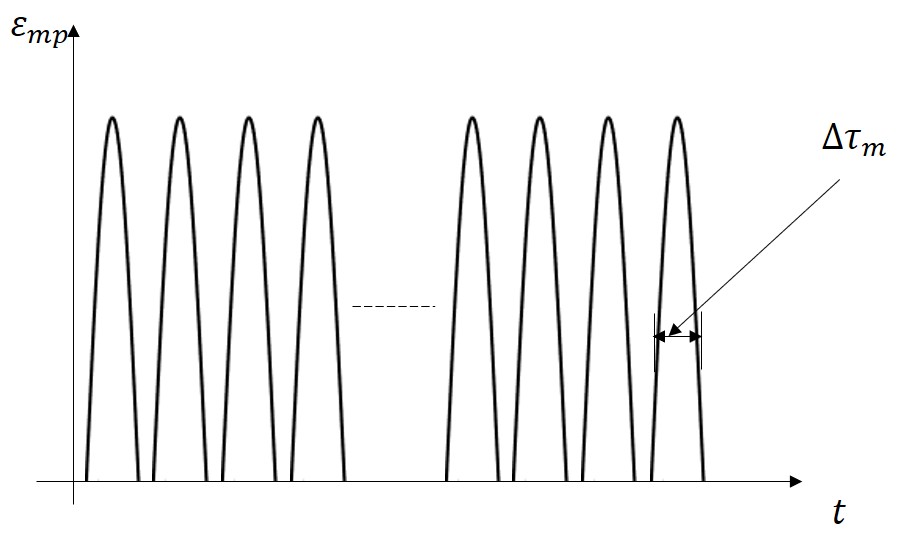}
	\caption{Schematic time dependence of the collision potential energy ${\varepsilon}_{mp}$ for collisions between molecules of types 1 and 2. Here, ${\varepsilon}_{mp}$ corresponds to ${\varepsilon}_{m12p}$ or ${\varepsilon}_{m21p}$ used in the theoretical formulation.}
\end{figure}

The average collision potential energy of each gas molecule can be expressed as follows:

\begin{equation}
	\begin{aligned}
		\overline{\varepsilon}_{1p} &= \overline{\epsilon}_{1p} \Delta\tau_1 Z_1 \\
		\overline{\varepsilon}_{2p} &= \overline{\epsilon}_{2p} \Delta\tau_2 Z_2 \\
		\overline{\varepsilon}_{m1p} &= \overline{\varepsilon}_{m11p} + \overline{\varepsilon}_{m12p}
		= \overline{\epsilon}_{1p} \Delta\tau_1 Z_{m1} + \overline{\epsilon}_{12p} \Delta\tau_{12} Z_{m12} \\
		\overline{\varepsilon}_{m2p} &= \overline{\varepsilon}_{m22p} + \overline{\varepsilon}_{m21p}
		= \overline{\epsilon}_{2p} \Delta\tau_2 Z_{m2} + \overline{\epsilon}_{21p} \Delta\tau_{21} Z_{m21}
	\end{aligned}\,,
	\label{AllCPotential}
\end{equation}

Here, $\overline{\epsilon}_{1p}$ and $\overline{\epsilon}_{2p}$ represent the average potential energy of a single collision between molecules of the same type, while $\Delta\tau_1$ and $\Delta\tau_2$ denote the typical duration of a collision between molecules of the same gas. $Z_1$ and $Z_2$ represent the number of collisions per unit time for molecules of the same gas, that is, the collision frequency.

$\overline{\varepsilon}_{m11p}$ represents the average collision potential energy between the type 1 molecules after mixing, while $\overline{\varepsilon}_{m12p}$ represents the average collision potential energy between the type 1 molecules and the type 2 molecules after mixing; $\overline{\varepsilon}_{m22p}$ represents the average collision potential energy between the type 2 molecules after mixing, and $\overline{\varepsilon}_{m21p}$ represents the average collision potential energy between the type 2 molecules and the type 1 molecules after mixing.

$\overline{\epsilon}_{12p}=\overline{\epsilon}_{21p}$ represent the average potential energy of a single collision between the two types of molecules after mixing, while $\Delta\tau_{12} = \Delta\tau_{21} = \Delta\tau_{m} $ denote the typical duration of a collision between the two types of gas molecules after mixing. 
$Z_{m1}$ and $Z_{m12}$, as well as $Z_{m2}$ and $Z_{m21}$, represent the number of collisions per unit time between molecules of the same gas after mixing, that is, the collision frequency.
In this equation, most of the terms are related to the temperature T. Except for $\Delta\tau_1$, $\Delta\tau_2$, $\Delta\tau_{12} = \Delta\tau_{21}$, which are related to the atomic properties and have no relation to the temperature.

\section{\label{sec:results}Results}

Combining the local-free-space perspective \cite{Guo:2020,Guo:2021} with the time-averaged collision potential energy, we formulate the canonical partition function using a local spatial contribution and collision-dependent energetic terms. The resulting entropy increment contains molecular and thermodynamic parameters, including molecular mass, effective radius, characteristic collision time, molecule number, volume, and temperature. Unlike Eq. \ref{MixingEntropyIncreaseSimple}, the resulting entropy increment is therefore not independent of molecular identity.

A central consequence is the identical-gas limit. When the two gases have the same molecular properties and are prepared under the same thermodynamic conditions, the parameters describing like and unlike collisions coincide and the mixing entropy increment becomes zero. For different gases, differences in molecular properties and collision characteristics generate a finite entropy increment. Thus, the transition from different to identical gases is determined by continuously varying physical parameters rather than by switching between distinguishable and indistinguishable counting rules.

\subsection{Mixing entropy increment--General}

According to the calculation method of the canonical ensemble\cite{StatMech:Books,StatMech:BooksAdv}, the free energy of gas molecules can be obtained as

\begin{equation}
	F = -Tk \ln\left( \sum_{n} e^{-E_n / kT} \right)\,,
	\label{FreeEnergyFormula1}
\end{equation}
Or

\begin{equation}
	F = -Tk \ln\left( \int e^{-\frac{E}{kT}} d\Gamma \right)
	= -Tk \ln\left( \frac{1}{(2\pi\hbar)^s} \int e^{-\frac{E(p,q)}{kT}} dp\,dq \right)\,,
	\label{FreeEnergyFormula2}
\end{equation}

Here, $E_n$ represents the discrete energy of all gas molecules, while $E$ represents the total energy after approximate continuous processing. $\Gamma$, $p$, and $q$ are phase space parameters, and $\hbar$ is the reduced Planck constant.

Because the distances between gas molecules are relatively large, their wave functions do not overlap. Therefore, they do not need to be regarded as identical particles in quantum mechanics. For a single type of gas, there is

\begin{equation}
	\begin{aligned}
		\sum_{n} e^{-E_{n}/T} = \left( \sum_{n} e^{-\varepsilon_{n}/kT} \right)^{N} \,,
		\label{FreeEnergyFormulaSumFactor}
	\end{aligned}	
\end{equation}
Here, $N$ represents the total number of molecules, $\varepsilon$ denotes the total energy of a single molecule, and $\varepsilon_{n}$ represents the discrete energy of a single molecule.

So we get
\begin{equation}
	\begin{aligned}
		F = -NkT \ln\left( \sum_{n} e^{-\varepsilon_{n}/kT} \right)	\,,
		\label{FreeEnergyFormula3}
	\end{aligned}	
\end{equation}

According to the definition, we have

\begin{equation}
	\varepsilon = \frac{\left(p_x^2 + p_y^2 + p_z^2\right)}{2m} + \varepsilon'\,,
	\label{EnergyFormula}
\end{equation}

Here, $p_x$, $p_y$, and $p_z$ represent the components of the molecular momentum, $m$ is the molecular mass, and $\varepsilon'$ represents other forms of energy apart from kinetic energy.
\begin{equation}
	\varepsilon' = \varepsilon_r + \varepsilon_v + \varepsilon_p\,,
	\label{EnergyFormulaOther}
\end{equation}
Here, $\varepsilon_r$ represents rotational energy, $\varepsilon_v$ represents vibrational energy, and $\varepsilon_p$ represents the potential energy of intermolecular collisions. $\varepsilon_r$ and $\varepsilon_v$ are the energies of the molecules themselves and have no influence on our calculation of the mixing entropy increment. They will be disregarded in the following text.

By integrating, we obtain:

\begin{equation}
	\begin{aligned}
		\sum_{n} e^{-\varepsilon_n/kT}
		&\cong \frac{1}{(2\pi\hbar)^3} \int \exp\left[-\varepsilon/kT\right] \mathrm{d}^3 p \,\mathrm{d}V \\
		&= \frac{1}{(2\pi\hbar)^3} \int \exp\left[
		-\left( \frac{p_x^2+p_y^2+p_z^2}{2m} + \varepsilon' \right)\big/kT
		\right] \mathrm{d}^3 p \,\mathrm{d}V \\
		&= \frac{1}{(2\pi\hbar)^3} \int \exp\left[
		-\left( \frac{p_x^2+p_y^2+p_z^2}{2m} \right)\big/kT
		\right] \mathrm{d}^3 p
		\int e^{-\frac{\varepsilon'}{kT}} \mathrm{d}V \\
		&\cong \frac{1}{(2\pi\hbar)^3} \int \exp\left[
		-\left( \frac{p_x^2+p_y^2+p_z^2}{2m} \right)\big/kT
		\right] \mathrm{d}^3 p
		\int e^{-\frac{\varepsilon_p}{kT}} \mathrm{d}V \\
		&\cong \left( \frac{mkT}{2\pi\hbar^2} \right)^{\frac{3}{2}}
		\left( \frac{V}{N} e^{-\overline{\varepsilon}_p/kT} \right)
	\end{aligned}\,,
	\label{IntegralFormula1}
\end{equation}

Here, $V$ represents the total volume of the gas, and as mentioned earlier, it has been assumed that the average free space volume for each molecule is $V/N$. $\overline{\varepsilon}_p$ is the average collisional potential energy.

Then it can be concluded that

\begin{equation}
	\begin{aligned}
		F &= -NkT \ln\left( \frac{V}{N} \left( \frac{mkT}{2\pi\hbar^2} \right)^{\frac{3}{2}} e^{-\overline{\varepsilon}_p/kT} \right) \\
		&= -NkT \ln\left( \frac{V}{N} \right) - NkT \ln\left[ \left( \frac{mkT}{2\pi\hbar^2} \right)^{\frac{3}{2}} \right] + N\overline{\varepsilon}_p
	\end{aligned}\,,
	\label{FreeEnergyFormula4}
\end{equation}

The entropy is:

\begin{equation}
	S = -\frac{\partial F}{\partial T}
	= Nk\ln\left(\frac{V}{N}\right)
	+ Nk\ln\left[\left(\frac{mkT}{2\pi\hbar^2}\right)^{\frac{3}{2}}\right]
	+ \frac{3}{2}Nk
	- N\frac{\partial\overline{\varepsilon}_p}{\partial T}\,,
	\label{EntropyFormula}
\end{equation}

The entropy of the two parts of gas when they exist separately is:
\begin{equation}
	\begin{aligned}
		S_1 &= -\frac{\partial F_1}{\partial T}
		= N_1 k \ln(V_1/N_1)
		+ N_1 k \ln\left( \left( \frac{m_1 k T}{2\pi \hbar^2} \right)^{\frac{3}{2}} \right)
		+ \frac{3}{2} N_1 k
		- N_1 \frac{\partial \overline{\varepsilon}_{1p}}{\partial T} \\[6pt]
		S_2 &= -\frac{\partial F_2}{\partial T}
		= N_2 k \ln(V_2/N_2)
		+ N_2 k \ln\left( \left( \frac{m_2 k T}{2\pi \hbar^2} \right)^{\frac{3}{2}} \right)
		+ \frac{3}{2} N_2 k
		- N_2 \frac{\partial \overline{\varepsilon}_{2p}}{\partial T}
	\end{aligned}\,,
	\label{EntropyFormulaSingle}
\end{equation}

Here, $V_1$ and $V_2$ represent the volumes of the two gases, $N_1$ and $N_2$ represent the total number of molecules of the two gases, $m_1$ and $m_2$ represent the molecular masses of the two gases, and $\overline{\varepsilon}_{1p}$ and $\overline{\varepsilon}_{2p}$ represent the average collision potential energy of each molecule of the two gases when they exist independently.

After the two parts of gas are mixed, the total volume and the total number of molecules satisfy:
\begin{equation}
	\begin{aligned}
		V &= V_1 + V_2 \\
		N &= N_1 + N_2
	\end{aligned}\,,
	\label{NVRelation}
\end{equation}

The free energy after the mixing of the two gases is:

\begin{equation}
	\begin{aligned}
		F_m
		&= -Tk \ln\left( \sum_{n} e^{-E_n/T} \right)
		= -Tk \ln\left[ \left( \sum_{i} e^{-\varepsilon_{m1,i}/kT} \right)^{N_1}
		\left( \sum_{j} e^{-\varepsilon_{m2,j}/kT} \right)^{N_2} \right] \\
		&= -Tk\left[ N_1 \ln\left( \sum_{i} e^{-\varepsilon_{m1,i}/kT} \right)
		+ N_2 \ln\left( \sum_{j} e^{-\varepsilon_{m2,j}/kT} \right) \right] \\
		&= -Tk\left[ N_1 \ln\left( \frac{V}{N} \left( \frac{m_1 kT}{2\pi\hbar^2} \right)^{\frac{3}{2}}
		e^{-\frac{\overline{\varepsilon}_{m1p}}{kT}} \right)
		+ N_2 \ln\left( \frac{V}{N} \left( \frac{m_2 kT}{2\pi\hbar^2} \right)^{\frac{3}{2}}
		e^{-\frac{\overline{\varepsilon}_{m2p}}{kT}} \right) \right] \\
		&= -N_1 Tk \ln\left( \frac{V}{N} \left( \frac{m_1 kT}{2\pi\hbar^2} \right)^{\frac{3}{2}} \right)
		+ N_1 \overline{\varepsilon}_{m1p}
		- N_2 Tk \ln\left( \frac{V}{N} \left( \frac{m_2 kT}{2\pi\hbar^2} \right)^{\frac{3}{2}} \right)
		+ N_2 \overline{\varepsilon}_{m2p}
	\end{aligned}\,,
	\label{FreeEnergyFormulaMixing}
\end{equation}
Here, $\varepsilon_{m1,i}$ and $\varepsilon_{m2,j}$ represent the total energy of each molecule in the mixture of the two gases, while $\overline{\varepsilon}_{m1p}$ and $\overline{\varepsilon}_{m2p}$ denote the average potential energy of each molecule in the mixture of the two gases. The rotational energy and vibrational energy have been disregarded in this context.

The entropy after mixing is:

\begin{equation}
	\begin{aligned}
		S_m = -\frac{\partial F_m}{\partial T}
		&= N_1 k \ln\left(\frac{V}{N}\right)
		+ N_1 k \ln\left( \left( \frac{m_1 k T}{2\pi \hbar^2} \right)^{\frac{3}{2}} \right)
		+ \frac{3}{2} N_1 k
		- N_1 \frac{\partial \overline{\varepsilon}_{m1p}}{\partial T} \\[6pt]
		&\quad + N_2 k \ln\left(\frac{V}{N}\right)
		+ N_2 k \ln\left( \left( \frac{m_2 k T}{2\pi \hbar^2} \right)^{\frac{3}{2}} \right)
		+ \frac{3}{2} N_2 k
		- N_2 \frac{\partial \overline{\varepsilon}_{m2p}}{\partial T}
	\end{aligned}\,,
	\label{EntropyFormulaMixing}
\end{equation}

The entropy increment of the mixed gas is:

\begin{equation}
	\begin{aligned}
		\Delta S &= S_m - (S_1 + S_2) \\
		&= -N_1 \left[ \frac{\partial \overline{\varepsilon}_{m1p}}{\partial T} - \frac{\partial \overline{\varepsilon}_{1p}}{\partial T} \right]
		- N_2 \left[ \frac{\partial \overline{\varepsilon}_{m2p}}{\partial T} - \frac{\partial \overline{\varepsilon}_{2p}}{\partial T} \right]
	\end{aligned}\,,
	\label{EntropyIncrease}
\end{equation}

\subsection{Mixing entropy increment--Simplified}

When gas molecules are separated, there is no potential energy between them. However, when the molecules collide, there is potential energy. And because the frequency of molecular collisions is extremely high, the potential energy cannot be ignored. We can take the time average of the collision potential energy as the average collision potential energy. So the average collision potential energy of each gas molecule can be expressed as follows:

Substituting Eq.\eqref{AllCPotential} into Eq.\eqref{EntropyIncrease}, we obtain the entropy increment of the mixed gas is:

\begin{equation}
	\begin{aligned}
		\Delta S &= S_m - (S_1 + S_2) \\
		&= -N_1\left[
		\frac{\partial\left(\overline{\epsilon}_{1p}\Delta\tau_1 Z_{m1} + \overline{\epsilon}_{12p}\Delta\tau_{12} Z_{m12}\right)}{\partial T}
		- \frac{\partial\left(\overline{\epsilon}_{1p}\Delta\tau_1 Z_1\right)}{\partial T}
		\right] \\
		&\quad -N_2\left[
		\frac{\partial\left(\overline{\epsilon}_{2p}\Delta\tau_2 Z_{m2} + \overline{\epsilon}_{21p}\Delta\tau_{21} Z_{m21}\right)}{\partial T}
		- \frac{\partial\left(\overline{\epsilon}_{2p}\Delta\tau_2 Z_2\right)}{\partial T}
		\right] \\
		&= -N_1\left[
		\Delta\tau_1 \frac{\partial\left(\overline{\epsilon}_{1p}Z_{m1} - \overline{\epsilon}_{1p}Z_1\right)}{\partial T}
		+ \Delta\tau_{12} \frac{\partial\left(\overline{\epsilon}_{12p}Z_{m12}\right)}{\partial T}
		\right] \\
		&\quad -N_2\left[
		\Delta\tau_2 \frac{\partial\left(\overline{\epsilon}_{2p}Z_{m2} - \overline{\epsilon}_{2p}Z_2\right)}{\partial T}
		+ \Delta\tau_{21} \frac{\partial\left(\overline{\epsilon}_{21p}Z_{m21}\right)}{\partial T}
		\right]
	\end{aligned}\,,
	\label{MixingEntropyIncrease-GG}
\end{equation}

For an ideal gas, the average collision potential energy before mixing can be expressed as:

\begin{equation}
	\overline{\varepsilon}_{1p} = \overline{\epsilon}_{1p}\Delta\tau_1 Z_1
	\approx \overline{\varepsilon}_{1k}\Delta\tau_1 Z_1
	= \left(\frac{3}{2}kT\right)\Delta\tau_1 Z_1\,,
	\label{CPotential1}
\end{equation}

\begin{equation}
	\overline{\varepsilon}_{2p} = \overline{\epsilon}_{2p}\Delta\tau_2 Z_2
	\approx \overline{\varepsilon}_{2k}\Delta\tau_2 Z_2
	= \left(\frac{3}{2}kT\right)\Delta\tau_2 Z_2\,,
	\label{CPotential2}
\end{equation}

Here, $\overline{\varepsilon}_{1k}$ and $\overline{\varepsilon}_{2k}$ represent the average translational kinetic energy of gas molecules.

\begin{equation}
	\begin{aligned}
		\overline{\epsilon}_{1p} &\approx \overline{\varepsilon}_{1k} = \frac{3}{2}kT \\[6pt]
		\overline{\epsilon}_{2p} &\approx \overline{\varepsilon}_{2k} = \frac{3}{2}kT
	\end{aligned}\,,
	\label{CPotentialMean}
\end{equation}

The number of collisions per unit time of the same type of gas molecules before mixing, that is, the collision frequency is:

\begin{equation}
	\begin{aligned}
		Z_1 &\approx \pi(2r_1)^2 \overline{u}_1 n_1
		= \pi(2r_1)^2 \sqrt{2}{\overline{v}_1}\,n
		= \pi(2r_1)^2 \sqrt{2}{\overline{v}_1}\,\frac{N_1+N_2}{V}\\[6pt]
		Z_2 &\approx \pi(2r_2)^2 \overline{u}_2 n_2
		= \pi(2r_2)^2 \sqrt{2}{\overline{v}_2}\,n
		= \pi(2r_2)^2 \sqrt{2}{\overline{v}_2}\,\frac{N_1+N_2}{V}
	\end{aligned}\,,
	\label{CollisionFrequency}
\end{equation}

Here, $r_1$ and $r_2$ represent the effective radii of the two types of gas molecules. $n_1 = N_1/V_1$ and $n_2 = N_2/V_2$ are the molecular number densities before the mixing. $n = (N_1 + N_2) / (V_1 + V_2) = (N_1 + N_2) / V = n_1 = n_2$ is the total molecular number density after the mixing. $\overline{u}_1 = \sqrt{2}{\overline{v}_1}$ and $\overline{u}_2 = \sqrt{2}{\overline{v}_2}$ are the average relative velocities between the same type of molecules, while $\overline{v}_1$ and $\overline{v}_2$ are the average velocities of the molecules.

\begin{equation}
	\begin{aligned}
		\overline{v}_1 &= \sqrt{\frac{8kT}{\pi m_1}} \\[6pt]
		\overline{v}_2 &= \sqrt{\frac{8kT}{\pi m_2}}
	\end{aligned}\,,
	\label{VelocityMean}
\end{equation}

After mixing, the average collision potential energy can be expressed as:

\begin{equation}
	\overline{\varepsilon}_{m1p} = \overline{\epsilon}_{1p}\Delta\tau_1 Z_{m1} + \overline{\epsilon}_{12p}\Delta\tau_{12} Z_{m12}
	= \left(\frac{3}{2}kT\right)\Delta\tau_1 Z_{m1} + \left(\frac{3}{2}kT\right)\Delta\tau_{12} Z_{m12}\,,
	\label{MixingCPotential1}
\end{equation}

\begin{equation}
	\overline{\varepsilon}_{m2p} = \overline{\epsilon}_{2p}\Delta\tau_2 Z_{m2} + \overline{\epsilon}_{21p}\Delta\tau_{21} Z_{m21}
	= \left(\frac{3}{2}kT\right)\Delta\tau_2 Z_{m2} + \left(\frac{3}{2}kT\right)\Delta\tau_{21} Z_{m21} \,,
	\label{MixingCPotential2}
\end{equation}
Among them
\begin{equation}
	\begin{aligned}
		\overline{\epsilon}_{12p} &\approx \big(\overline{\varepsilon}_{1k} + \overline{\varepsilon}_{2k}\big)/2 = \frac{3}{2}kT \\[6pt]
		\overline{\epsilon}_{21p} &\approx \big(\overline{\varepsilon}_{1k} + \overline{\varepsilon}_{2k}\big)/2 = \frac{3}{2}kT
	\end{aligned}\,,
	\label{CPotentialSingle}
\end{equation}
\begin{equation}
	\begin{aligned}
		Z_{m1} &\approx \pi(2r_1)^2 \overline{u}_1 n_{m1}
		= \pi(2r_1)^2 \sqrt{2}{\overline{v}_1}\, n_{m1}
		= \pi(2r_1)^2 \sqrt{2}{\overline{v}_1}\,\frac{N_1}{V}\\[6pt]
		Z_{m2} &\approx \pi(2r_2)^2 \overline{u}_2 n_{m2}
		= \pi(2r_2)^2 \sqrt{2}{\overline{v}_2}\, n_{m2}
		= \pi(2r_2)^2 \sqrt{2}{\overline{v}_2}\,\frac{N_2}{V}
	\end{aligned}\,,
	\label{CollisionFrequencyMixing}
\end{equation}
\begin{equation}
	\begin{aligned}
		Z_{m12} &\approx \pi(r_1+r_2)^2 \overline{u}_{12} n_{m2}
		= \pi(r_1+r_2)^2 \sqrt{\overline{v}_1^2 + \overline{v}_2^2}\,\frac{N_2}{V}\\[6pt]
		Z_{m21} &\approx \pi(r_1+r_2)^2 \overline{u}_{21} n_{m1}
		= \pi(r_1+r_2)^2 \sqrt{\overline{v}_1^2 + \overline{v}_2^2}\,\frac{N_1}{V}
	\end{aligned}\,,
	\label{CollisionFrequencyMixingInter}
\end{equation}

Here, $n_{m1} = N_1/V$ and $n_{m2} = N_2/V$ represent the density of molecules after the mixing, and $n = n_{m1} + n_{m2}$. $\overline{u}_{12} = \overline{u}_{21} = \sqrt{\overline{v}_1^2 + \overline{v}_2^2}$ is the average relative speed between the molecules of the two gases after the mixing.

Assume

\begin{equation}
	\zeta = \left( \frac{3}{2} k \right) \sqrt{\frac{8k}{\pi}}\,,
	\label{Constant1}
\end{equation}

Then
\begin{equation}
	\begin{aligned}
		\overline{\varepsilon}_{1p}
		&= \left(\frac{3}{2}kT\right)\Delta\tau_1 Z_1
		= \left(\frac{3}{2}kT\right)\Delta\tau_1 \pi(2r_1)^2 \overline{u}_1 n
		= \left(\frac{3}{2}kT\right)\Delta\tau_1 \pi(2r_1)^2 \sqrt{2\overline{v}_1}\,n \\[6pt]
		&= \left(\frac{3}{2}kT\right)\Delta\tau_1 \pi(2r_1)^2 \sqrt{2\overline{v}_1}\,\frac{N_1+N_2}{V}
		= \left(\frac{3}{2}kT\right)\Delta\tau_1 \pi(2r_1)^2 \sqrt{2}\sqrt{\frac{8kT}{\pi m_1}}\,\frac{N_1+N_2}{V} \\[6pt]
		&= \zeta T^{3/2}\Delta\tau_1 (2r_1)^2 \sqrt{2}\sqrt{\frac{1}{m_1}}\,\frac{N_1+N_2}{V}
	\end{aligned}\,,
	\label{CPotential1-X}
\end{equation}

\begin{equation}
	\begin{aligned}
		\overline{\varepsilon}_{2p}
		&= \left(\frac{3}{2}kT\right)\Delta\tau_2 Z_2
		= \left(\frac{3}{2}kT\right)\Delta\tau_2 \pi(2r_2)^2 \overline{u}_2 n
		= \left(\frac{3}{2}kT\right)\Delta\tau_2 \pi(2r_2)^2 \sqrt{2\overline{v}_2}\,n \\[6pt]
		&= \left(\frac{3}{2}kT\right)\Delta\tau_2 \pi(2r_2)^2 \sqrt{2\overline{v}_2}\,\frac{N_1+N_2}{V}
		= \left(\frac{3}{2}kT\right)\Delta\tau_2 \pi(2r_2)^2 \sqrt{2}\sqrt{\frac{8kT}{\pi m_2}}\,\frac{N_1+N_2}{V} \\[6pt]
		&= \zeta T^{3/2}\Delta\tau_2 (2r_2)^2 \sqrt{2}\sqrt{\frac{1}{m_2}}\,\left|\frac{N_1+N_2}{V}\right|
	\end{aligned}\,,
	\label{CPotential2-X}
\end{equation}

\begin{equation}
	\begin{aligned}
		\overline{\varepsilon}_{m1p}
		&= \left(\frac{3}{2}kT\right)\Delta\tau_1\pi(2r_1)^2\overline{u}_1 n_{m1}
		+ \left(\frac{3}{2}kT\right)\Delta\tau_{12}\pi(r_1+r_2)^2\overline{u}_{12} n_{m2} \\[6pt]
		&= \left(\frac{3}{2}kT\right)\Delta\tau_1\pi(2r_1)^2\sqrt{2\overline{v}_1}\frac{N_1}{V}
		+ \left(\frac{3}{2}kT\right)\Delta\tau_{12}\pi(r_1+r_2)^2\sqrt{\overline{v}_1^2+\overline{v}_2^2}\frac{N_2}{V} \\[6pt]
		&= \left(\frac{3}{2}kT\right)\Delta\tau_1\pi(2r_1)^2\sqrt{2}\sqrt{\frac{8kT}{\pi m_1}}\frac{N_1}{V}
		+ \left(\frac{3}{2}kT\right)\Delta\tau_{12}\pi(r_1+r_2)^2\sqrt{\frac{8kT}{\pi}}\sqrt{\frac{1}{m_1}+\frac{1}{m_2}}\frac{N_2}{V} \\[6pt]
		&= \zeta T^{3/2}\left(
		\Delta\tau_1(2r_1)^2\sqrt{2}\sqrt{\frac{1}{m_1}}\frac{N_1}{V}
		+ \Delta\tau_{12}(r_1+r_2)^2\sqrt{\frac{1}{m_1}+\frac{1}{m_2}}\frac{N_2}{V}
		\right)
	\end{aligned}\,,
	\label{MixingCPotential1-X}
\end{equation}

\begin{equation}
	\begin{aligned}
		\overline{\varepsilon}_{m2p}
		&= \left(\frac{3}{2}kT\right)\Delta\tau_2\pi(2r_2)^2\overline{u}_2 n_{m2}
		+ \left(\frac{3}{2}kT\right)\Delta\tau_{21}\pi(r_1+r_2)^2\overline{u}_{21} n_{m1} \\[6pt]
		&= \left(\frac{3}{2}kT\right)\Delta\tau_2\pi(2r_2)^2\sqrt{2\overline{v}_2}\frac{N_2}{V}
		+ \left(\frac{3}{2}kT\right)\Delta\tau_{21}\pi(r_1+r_2)^2\sqrt{\overline{v}_1^2+\overline{v}_2^2}\frac{N_1}{V} \\[6pt]
		&= \left(\frac{3}{2}kT\right)\Delta\tau_2\pi(2r_2)^2\sqrt{2}\sqrt{\frac{8kT}{\pi m_2}}\frac{N_2}{V}
		+ \left(\frac{3}{2}kT\right)\Delta\tau_{21}\pi(r_1+r_2)^2\sqrt{\frac{8kT}{\pi}}\sqrt{\frac{1}{m_1}+\frac{1}{m_2}}\frac{N_1}{V} \\[6pt]
		&= \zeta T^{3/2}\left(
		\Delta\tau_2(2r_2)^2\sqrt{2}\sqrt{\frac{1}{m_2}}\frac{N_2}{V}
		+ \Delta\tau_{21}(r_1+r_2)^2\sqrt{\frac{1}{m_1}+\frac{1}{m_2}}\frac{N_1}{V}
		\right)
	\end{aligned}\,,
	\label{MixingCPotential2-X}
\end{equation}

The entropy increment of the mixed gas is:

\begin{equation}
	\begin{aligned}
		\Delta S &= S_m - (S_1 + S_2) \\
		&= -N_1\left[\frac{\partial \overline{\varepsilon}_{m1p}}{\partial T} - \frac{\partial \overline{\varepsilon}_{1p}}{\partial T}\right]
		- N_2\left[\frac{\partial \overline{\varepsilon}_{m2p}}{\partial T} - \frac{\partial \overline{\varepsilon}_{2p}}{\partial T}\right] \\
		&\approx -N_1\left[
		\zeta \frac{3}{2}T^{1/2}\left(
		\Delta\tau_1(2r_1)^2\sqrt{2}\sqrt{\frac{1}{m_1}}\frac{N_1}{V}
		+ \Delta\tau_{12}(r_1+r_2)^2\sqrt{\frac{1}{m_1}+\frac{1}{m_2}}\frac{N_2}{V}
		\right)
		\right. \\
		&\quad\left.
		- \zeta \frac{3}{2}T^{1/2}\Delta\tau_1(2r_1)^2\sqrt{2}\sqrt{\frac{1}{m_1}}\frac{N_1+N_2}{V}
		\right] \\
		&\quad -N_2\left[
		\zeta \frac{3}{2}T^{1/2}\left(
		\Delta\tau_2(2r_2)^2\sqrt{2}\sqrt{\frac{1}{m_2}}\frac{N_2}{V}
		+ \Delta\tau_{21}(r_1+r_2)^2\sqrt{\frac{1}{m_1}+\frac{1}{m_2}}\frac{N_1}{V}
		\right)
		\right. \\
		&\quad\left.
		- \zeta \frac{3}{2}T^{1/2}\Delta\tau_2(2r_2)^2\sqrt{2}\sqrt{\frac{1}{m_2}}\frac{N_1+N_2}{V}
		\right] \\
		&= -N_1\zeta\frac{3}{2}T^{1/2}\left[
		\Delta\tau_{12}(r_1+r_2)^2\sqrt{\frac{1}{m_1}+\frac{1}{m_2}}\frac{N_2}{V}
		- \Delta\tau_1(2r_1)^2\sqrt{2}\sqrt{\frac{1}{m_1}}\frac{N_2}{V}
		\right] \\
		&\quad -N_2\zeta\frac{3}{2}T^{1/2}\left[
		\Delta\tau_{21}(r_1+r_2)^2\sqrt{\frac{1}{m_1}+\frac{1}{m_2}}\frac{N_1}{V}
		- \Delta\tau_2(2r_2)^2\sqrt{2}\sqrt{\frac{1}{m_2}}\frac{N_1}{V}
		\right] \\
		&= \frac{N_1N_2}{V}\zeta\frac{3}{2}T^{1/2}\left[
		\sqrt{2}\Delta\tau_1(2r_1)^2\sqrt{\frac{1}{m_1}}
		+ \sqrt{2}\Delta\tau_2(2r_2)^2\sqrt{\frac{1}{m_2}}
		- 2\Delta\tau_{m}(r_1+r_2)^2\sqrt{\frac{1}{m_1}+\frac{1}{m_2}}
		\right]
	\end{aligned}\,,
	\label{MixingEntropyIncrease}
\end{equation}
Here, the relationship $\Delta\tau_{12} = \Delta\tau_{21} = \Delta\tau_{m}$ has been applied.

That is to say:
\begin{equation}
	\Delta S \approx \frac{N_1N_2}{V}\zeta \frac{3}{2}T^{1/2}
	\left[
	\sqrt{2}\Delta\tau_1(2r_1)^2\sqrt{\frac{1}{m_1}}
	+ \sqrt{2}\Delta\tau_2(2r_2)^2\sqrt{\frac{1}{m_2}}
	- 2\Delta\tau_{m}(r_1+r_2)^2\sqrt{\frac{1}{m_1}+\frac{1}{m_2}}
	\right]\,,
	\label{MixingEntropyIncreaseSimplified}
\end{equation}

This equation indicates that the entropy increment of gas mixing is related to parameters such as the number of molecules, volume, temperature, effective radius of molecules, molecular mass, and typical collision duration. It is not a fixed constant; when the types of the two gas molecules are the same, the entropy increment is precisely 0.

\section{\label{sec:discussion}Discussion}
	
The Gibbs paradox has generated several conceptually distinct lines of resolution. One influential class of approaches invokes quantum mechanics. Von Neumann \cite{Von Neumann:1932} related particle similarity to the overlap between quantum states, allowing the mixing entropy to vary continuously as the states become more similar; related arguments have subsequently been developed \cite{Unnikrishnan:2018}. These treatments establish an important connection between distinguishability and quantum states, but they address the problem through quantum entropy or quantum-state structure rather than through the classical thermodynamic mechanism considered here.

A second line of thought treats entropy, or the entropy of mixing, as dependent on the observer's ability to distinguish or manipulate components. Van Kampen \cite{VanKampen:1984} emphasized the operational conditions under which two gases can be distinguished, an approach further developed in related work \cite{Allahverdyan-Balian-Nieuwenhuizen:2004,Allahverdyan-Nieuwenhuizen:2006,Yadin-Morris-Adesso:2021}. Jaynes similarly connected entropy to the information available to an observer \cite{Jaynes:1992}, and more recent discussions have related distinguishability to observational technology and information acquisition \cite{Ntantis-Xezonakis:2025}; related perspectives have also been discussed by Saunders \cite{Saunders:2018}. Lin's information-theoretic formulation instead assigns zero thermodynamic entropy increment to ideal mixing and attributes the mixing effect to information loss, with an information entropy that depends on molecular similarity \cite{LinSK:2008}. These approaches clarify the roles of information and operational distinguishability. The present framework addresses a different question: whether a property-dependent mixing entropy can be obtained as an objective thermodynamic quantity within a classical molecular description.

Another approach distinguishes thermodynamic entropy from statistical entropy. Dieks and coworkers \cite{Dieks:2010,Versteegh-Dieks:2011,Dieks:2018} argued that mixing identical gases produces no thermodynamic entropy change even though individual particles may remain distinguishable in a microscopic description. In such a view, statistical state counting and thermodynamic entropy need not coincide in the conventional manner. Our formulation instead seeks to retain a direct connection between a classical statistical-mechanical construction and the thermodynamic entropy change.

Other studies modify the mathematical or probabilistic description of mixing. Maslov \cite{Maslov:2011} introduced a number-theoretic counting framework with a continuous fractional dimension $d$, making entropy continuous as $d$ varies, although this construction introduces additional structures such as clusters or dimers. Paillusson \cite{Paillusson:2023} formulated the problem in terms of composition probability distributions and finite-size fluctuations, providing a continuous description as molecular similarity changes; the resulting free-space contribution includes an $\mathcal{O}(\ln N)$ term and therefore raises a finite-size question concerning strict extensivity. Swendsen \cite{Swendsen:2018} instead emphasized probabilities of particle exchange among multiple systems and showed that quantum mechanics need not be essential to the resolution, although the accessible spatial integration remains global rather than local.

The local-free-space perspective is especially relevant to the present work. Previous studies proposed that the instantaneous free volume of each molecule should be local and concluded that mixing need not increase entropy \cite{Guo:2020,Guo:2021}. This construction provides a route to extensivity but, by itself, does not readily account for entropy production during irreversible mixing of distinct gases. Our extension is to retain the time-averaged potential energy associated with molecular collisions. The combination of local spatial accessibility and collision energetics introduces explicit molecular parameters into the entropy increment of mixing and thereby distinguishes different gases with a continuous change of entropy increment.

The framework has two main implications. First, it provides a classical mechanism by which the mixing entropy can depend on molecular identity through measurable or, in principle, estimable physical quantities such as mass, effective radius, and collision time. Second, the identical-gas limit follows from equality of these molecular and collision parameters, so the entropy increment approaches zero without introducing quantum indistinguishability as an additional thermodynamic postulate. These features distinguish the present formulation from interpretations based primarily on information, observer-dependent distinguishability, or other routes mentioned above.

Direct experimental determination of the entropy increment attributable solely to gas mixing remains challenging. A possible thermodynamic route was discussed by Ihnatovych \cite{Ihnatovych:2013}, who considered reconstructing the entropies of pure and mixed substances from absolute zero using heat capacities, phase-transition temperatures, enthalpy changes, and related thermodynamic data. In such a construction, identical gases yield zero mixing entropy, whereas mixtures of different substances can yield a nonzero value that depends on component and mixture properties. Comparison between independently measurable thermodynamic entropy changes and predictions of the present formulation could therefore provide a route for testing the model.

\section{\label{sec:conclusion}Conclusion}

By combining instantaneous local-free-space perspective with time-averaged collision potential energy in a canonical ensemble framework, we obtain a classical mixing entropy that depends explicitly on molecular properties and vanishes in the identical-gas limit. The formulation therefore provides a mechanistic route to the Gibbs paradox without using quantum indistinguishability as the physical basis of the entropy increment. More broadly, it establishes a testable connection between molecular properties and macroscopic entropy increment during mixing.

The physical picture may also motivate extensions beyond dilute gases. Local free volume and intermolecular interactions are central to diffusion and mixing in condensed phases, suggesting that related ideas could be explored for liquid mixtures. Such an extension would require an explicit treatment of stronger and more complex intermolecular interactions, phase behavior, and nonideal effects; accordingly, phenomena such as ink dispersion in water or oil-water immiscibility should be regarded as potential applications to be tested rather than as consequences already established by the present gas-phase derivation.
	
\begin{acknowledgments}
This work was supported by the Scientific Research Starting Project for Advanced Imported Talents of Wuyi University (YJ202011).
We also thank our anonymous referees for their very helpful comments and suggestions.
\end{acknowledgments}
	


\begin{thebibliography}{99}
		
\bibitem{StatMech:Books}
Z.-C. Wang, 
\textit{Thermodynamics and Statistical Physics}, 5th ed., Higher Education Press, Beijing, 2013.\\
D. Kondepudi and I. Prigogine, 
\textit{Modern Thermodynamics: From Heat Engines to Dissipative Structures}; 2nd ed., John Wiley \& Sons, Ltd, 2015.\\
R. K. Pathria and P. D. Beale, 
\textit{Statistical Mechanics}; 3rd ed., Elsevier Ltd., 2011.\\
Carl S. Helrich,
\textit{Modern Thermodynamics with Statistical Mechanics};
Springer-Verlag Berlin Heidelberg, 2009.\\
and many other textbooks.

\bibitem{StatMech:BooksAdv}
L. D. Landau and E. M. Lifshitz, \textit{Statistical Physics, Part 1}, 3rd ed., Course of Theoretical Physics, Vol. 5, Pergamon Press, Oxford, 1980.

\bibitem{Darrigol:2018}
O. Darrigol,  
The Gibbs paradox: Early history and solutions, 
Entropy 2018, 20, 443.

\bibitem{LinSK-webpage:2009}  
A web URL collecting many literatures about Gibbs padadox: https://www.mdpi.org/lin/entropy/gibbs-paradox.htm, 
Compiled by Shu-Kun Lin. Last change: 7 November 2009.

\bibitem{Gibbs:1879}
J. W. Gibbs,  
\textit{On the equilibrium of heterogeneous substances}; 
Trans. Conn. Acad. Arts Sci. 1879, III, 108–248.

\bibitem{Gibbs:1902}
J. W. Gibbs,  
\textit{Elementary Principles in Statistical Mechanics};
Yale University Press: New Haven, CT, USA, 1902.

\bibitem{Casper-Freier:1973}
B. Casper, and S. Freier,  
‘Gibbs Paradox’ paradox, 
Am. J. Phys. 1972, 41, 509–511.

\bibitem{VanKampen:1984}
N. G. van Kampen,  
The Gibbs’ paradox. In \textit{Proceedings of the Essays in Theoretical Physics: in Honor of Dirk ter Haar}; Parry, W.E., Ed.; Pergamon: Oxford, UK, 1984.

\bibitem{Zheng:1987} 
J.-R. Zheng,  Thermodynamic solution of the Gibbs paradox, 
College Physics. 1987, 1(3), 1-1.


\bibitem{Jaynes:1992}
E. T. Jaynes,  
The Gibbs’ paradox. In \textit{Proceedings of the Maximum Entropy and Bayesian Methods}; Smith, C., Erickson, G., Neudorfer, P., Eds.; Kluwer Academic: Norwell, MA, USA, 1992; pp. 1–22.

\bibitem{LinSK:2008} 
S.-K. Lin, 
Gibbs Paradox and the Concepts of Information, Symmetry, Similarity and Their
Relationship, 
Entropy 10, 1-5 (2008). arXiv:0803.2571.


\bibitem{Dieks:2010}
D. Dieks, 
The Gibbs Paradox Revisited. In \textit{Explanation, Prediction, and Confirmation}, 
edited by Dennis Dieks et al., Springer, 2010, pp. 367–377.

\bibitem{Versteegh-Dieks:2011} 
M. A. Versteegh, and D. Dieks,  
The Gibbs paradox and the distinguishability of identical particles, 
American Journal of Physics 79, 741-746 (2011).

\bibitem{Dieks:2018}
D. Dieks, 
The Gibbs Paradox and Particle Individuality,  
Entropy. 2018; 20(6):466. https://doi.org/10.3390/e20060466

\bibitem{Maslov:2011} 
V. P. Maslov,
Mathematical Solution of the Gibbs Paradox,
ISSN 0001-4346, Mathematical Notes, 2011, Vol. 89, No. 2, pp. 266–276.

\bibitem{Ge-Qian:2011} 
H. Ge and H. Qian, 
Maximum Entropy Principle, Equal Probability a Priori and Gibbs Paradox,
arXiv:1105.4118.

\bibitem{Ihnatovych:2013}
V. Ihnatovych,  
Study of the possibility of eliminating the Gibbs paradox within the framework of classical thermodynamics,
arXiv:1306.5737.

\bibitem{Saunders:2018}
S. Saunders,
The Gibbs Paradox,
Entropy 2018, 20(8), 552.

\bibitem{Swendsen:2018}
R. H. Swendsen, 
Probability, Entropy, and Gibbs’ Paradox(es), 
Entropy. 2018; 20(6):450. https://doi.org/10.3390/e20060450

\bibitem{Unnikrishnan:2018}
C. S. Unnikrishnan,
The Gibbs Paradox and the Physical Criteria for the Indistinguishability of Identical,
arXiv:1811.03967

\bibitem{Guo:2020}
Quanmin Guo, 
Redefining the Phase Space for Ideal Gas Systems Resolves the Gibbs Paradox, 
arXiv:2004.12228.

\bibitem{Guo:2021}
Quanmin Guo, 
The Gibbs paradox, 
arXiv:2106.05868.

\bibitem{Paillusson:2023}
Paillusson F. The “Real” Gibbs Paradox and a Composition-Based Resolution. Entropy. 2023; 25(6):833. https://doi.org/10.3390/e25060833

\bibitem{Tao:2025}
Y. Tao,  
Gibbs paradox and thermodynamics of colloids,
Physics Letters A, 547, 130531 (2025). https://doi.org/10.1016/j.physleta.2025.130531

\bibitem{Ntantis-Xezonakis:2025}
E. L. Ntantis, and V. Xezonakis, 
Revisiting Gibbs paradox: Resolving misconceptions and addressing challenges,
International Journal of Innovative Research and Scientific Studies, 8(3), pages: 4411-4417, (2025).

\bibitem{Einstein:1924}
A. Einstein, 
Quantentheorie des einatomigen idealen Gases. Berl. Ber. 261–267, 1924.

\bibitem{Von Neumann:1932}
John Von Neumann, 
\textit{Mathematische Grundlagen der Quantenmechanik}; Springer: Berlin, Germany, 1932.
John von Neumann, translated from the German by Robert T. Beyer, edited by Nicholas A. Wheeler
\textit{Mathematical Foundations of Quantum Mechanics}; New Edition, Princeton University Press, 2018.

\bibitem{Bazarov:1975}
I.P. Bazarov,  
Einstein's paradox and a new mixing paradox. 
Soviet Physics Journal 18, 626–629 (1975). 
https://doi.org/10.1007/BF00893992

\bibitem{Allahverdyan-Balian-Nieuwenhuizen:2004}
A. E. Allahverdyan, R. Balian and Th. M. Nieuwenhuizen, 
Maximal work extraction from finite quantum systems,
Europhysics Letters 67, 565, 2004.
https://doi.org/10.1209/epl/i2004-10101-2

\bibitem{Allahverdyan-Nieuwenhuizen:2006}
A. E. Allahverdyan, Th. M. Nieuwenhuizen,
Explanation of the Gibbs paradox within the framework of quantum thermodynamics,
Phys. Rev. E 73, 066119, 2006. 
https://doi.org/10.1103/PhysRevE.73.066119


\bibitem{Yadin-Morris-Adesso:2021}
B. Yadin, B. Morris, \& G. Adesso, 
Mixing indistinguishable systems leads to a quantum Gibbs paradox. Nature Communications, 12(1), 1471 (2021). 
https://doi.org/10.1038/s41467-021-21620-7
		
\end{thebibliography}


\end{document}